\documentclass{article}

\usepackage{arxiv}

\usepackage[utf8]{inputenc} 
\usepackage[T1]{fontenc}    
\usepackage{url}            
\usepackage{booktabs}       
\usepackage{amsfonts}       
\usepackage{nicefrac}       
\usepackage{microtype}      
\usepackage{lipsum}		
\usepackage{graphicx}
\usepackage[square,sort,comma,numbers]{natbib}
\usepackage{doi}
\usepackage{indentfirst}

\title{\emph{FAIR} begins at home: Implementing \emph{FAIR} via the Community of Data Driven Insights}

\author{ \href{https://orcid.org/0000-0002-9994-1462}{\includegraphics[scale=0.06]{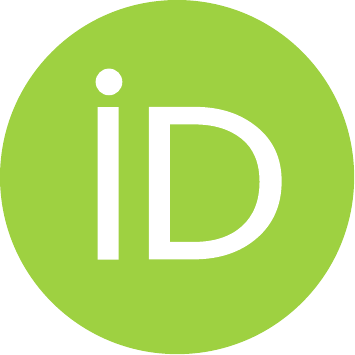}\hspace{1mm}Carlos Utrilla Guerrero}\thanks{This is a pre-print of the CDDI, a research project presented at the FDO2022 conference, see \citep{home}. Corresponding author: Carlos Utrilla Guerrero, \textit{Now working at TU Delft University: c.utrillaguerrero@tudelft.nl.}\hspace{1mm}}\ \\
	Institute of Data Science (IDS)\\
	Maastricht University (UM)\\
	Paul-Henri Spaaklaan 1  \\
	\And
    {\hspace{1mm}Maria Vivas Romero} \\
    UB Research Support and Development \\
    University Library (UL) \\
	College van Bestuur \\
	\And
	{\hspace{1mm}Marc Dolman} \\
    Maastricht University Office \\
    Maastricht University (UM) \\
	Minderbroedersberg 4-6 \\
	\And
	\href{https://orcid.org/0000-0003-4727-9435}{\includegraphics[scale=0.06]{orcid.pdf}\hspace{1mm}Michel Dumontier} \\
	Institute of Data Science (IDS)\\
	Maastricht University (UM)\\
	Paul-Henri Spaaklaan 1  \\
}

\renewcommand{\undertitle}{Technical Report}

\hypersetup{
pdftitle={FAIR begins at home},
pdfsubject={cs.DL},
pdfauthor={Carlos Utrilla Guerrero},
pdfkeywords={FAIR principles, Data Science, Research data management services},
}

\begin{document}
\maketitle

\begin{abstract}
Arguments for the FAIR principles have mostly been based on appeals to values. However, the work of onboarding diverse researchers to make efficient and effective implementations of FAIR requires different appeals. In our recent effort to transform the institution into a FAIR University by 2025, here we report on the experiences of the Community of Data Driven Insights (CDDI). We describe these experiences from the perspectives of a data steward in social sciences and a data scientist, both of whom have been working in parallel to provide research data management and data science support to different research groups. We initially identified 5 challenges for FAIR implementation. These perspectives show the complex dimensions of FAIR implementation to researchers across disciplines in a single university.
\end{abstract}

\vspace{5mm} 

\setlength{\parindent}{1em}
\section{Introduction}

The FAIR principles\citep{wilkinson2016fair} are well-known forward-thinking guidelines that improve the infrastructure for the reuse of scholarly digital objects to ensure transparency, reproducibility, and reusability. \footnote{FAIR-based thinking is gaining traction within academia, as the principles have been adopted by the \href{https://ec.europa.eu/commission/presscorner/detail/en/STATEMENT_16_2967}{G20}, European Open Science Cloud Infrastructure \href{https://eosc-portal.eu/sites/default/files/eosc_declaration.pdf}{EOSC}, the EU’s Framework Programmes for funding  - \href{https://ec.europa.eu/research/participants/data/ref/h2020/grants_manual/hi/oa_pilot/h2020-hi-oa-data-mgt_en.pdf}{Horizon 2020}, the Dutch Research Organization \href{https://www.nwo.nl/en/research-data-management}{NWO}, and the U.S. National Institutes of Health - \href{https://grants.nih.gov/grants/guide/notice-files/NOT-OD-21-013.html}{NIH} as well as research institutions, other funding agencies, private companies and journals}. The current practice of FAIR implementation however has a number of significant barriers that prevent its global adoption, which has been subject of discussion lately: among others, the limited understanding and availability of mature FAIR related technology, plausible siloed data mentality of the relevant stakeholders and lack of secure long-term return on investment. \citep{wise2019implementation}

In this commentary, we report from the front lines of FAIR implementation within a single institutional setting. By disclosing the progress and challenges of implementing FAIR, we hope to shed light on the process in a way that might be useful for other institutions in Europe and elsewhere.

\subsection{The FAIR principles, its benefits and limitations}

Science is opening up more, not only in terms of data but also in terms of other digital objects such as computer code, software and workflows. Sharing all kinds of digital objects is important because it yields benefits such as new collaborations, increases confidence in findings and generates goodwill among research communities. But most importantly, reusing these digital objects in the scientific community fosters better science. The necessity of improving science (data organization, storage efficiency and effectiveness) become evidence. The FAIR principles emerged as a potential solution to help transform science.

The FAIR principles have shown already significant positive effects to the economy \footnote{Not having FAIR research data costs the European economy around euro 10.2bn/yr. See more info: \href{https://op.europa.eu/en/publication-detail/-/publication/d375368c-1a0a-11e9-8d04-01aa75ed71a1}{(Cost-benefit analysis for FAIR research data)}} as well as the ability to provide benefits to a variety of stakeholders, including researchers wishing to share and reuse data, scientific publishers or funding agencies for data stewardship, software providers for data storage, processing and analysis, and the data science community using data to advance scientific discovery. Implementing FAIR helps researchers to guarantee not only discoverability and visibility of researchers’ outputs, but also – and more fundamentally – improve the credibility and veracity of their knowledge claims. FAIR way of research data management \footnote{The term ‘research data management’ covers a set of activities related to how researchers save, organize, storage and describe the materials they work with over the course of a research project} is seen as a means to secure the findability and reusability of these digital resources.

The wide range of scientific, cultural and organizational characteristics however, challenges the implementation of FAIR on the ground. The FAIR principles implementation is probably in its very early stage, involving a continuous learning process from all stakeholders, but it is imperative that researchers increasingly endorse these principles for a good data management and sharing: digital resources should be easily \textit{Findable} and \textit{Accessible} in an user-friendly and machine-readable manner, \textit{Interoperable} to smoothly connect and understand with ‘others’ (people and machines) and \textit{Reusable} using widely community standards.

Despite showing potential benefits that go beyond researchers' scientific soundness of research practices and project outcomes \citep{leveque2013top}, many of them are still far from these principles into practice\citep{watson2022many}. We found that a journey headed towards good research data management (RDM) captained by FAIR guidelines can be sometimes fraught - personally, financially and technically. Personally because often it takes so long with barely perceptible and tangible benefits. Financially because researchers lack the budget for implementing FAIR, and technically due to the absence of mature-status technology, proliferation of digital tools, and lack of agreeing standards, guidelines and procedures.

\subsection{The Community for Data Driven Insights (CDDI) for FAIR practice}

Maastricht University (UM) aims to be entirely FAIR by 2025. The CDDI was created as an interfaculty resource representing scientific stakeholders across the six university faculties. This inter-departmental initiative aims to turn all digital objects within UM into FAIR digital objects. As the CDDI was founded, international initiatives such as \href{https://www.go-fair.org/}{GO FAIR}, \href{https://fairdo.org/}{FAIR Digital Objects}, \href{https://www.rd-alliance.org/}{Research Data Alliance}, and \href{https://eosc-portal.eu/}{EOSC} were the norm, but national initiatives (e.g. \href{https://www.nwo.nl/en/news/digital-competence-centers-knowledge-institutions-forging-ahead}{Local Digital Competence Centers - DCC} and \href{https://data.4tu.nl/info/en/}{4TU.ResearchData}) were also being set up in order to understand what does mean implementing the FAIR data principles in universities in reality. These initiatives aim to increase universities’ capability in research data management and foster alignment across the different sectors. 

We started this reflection as a data scientist and data steward, respectively, employed by CDDI which was originally founded in 2018 to promote and develop UM as a FAIR university. All UM research data service providers are joint together in the CDDI to support researchers and research groups with all aspects concerning research data management\href{https://library.maastrichtuniversity.nl/research/rdm/rdm-and-fair/}{(RDM)}. We reached out to researchers via CDDI showcases across the UM\footnote{An example of CDDI Showcase: \url{https://library.maastrichtuniversity.nl/research/rdm/rdm-and-fair/lawgex/}} to help them understand how they can engage with the FAIR guidelines per each stage of the research lifecycle, and actively advise them to make decisions about their implementation choice per principle. These inter-disciplinary and inter-faculty collaborations were chosen with a bottom-up approach, they obtained our support as data steward and data scientist to advance in the FAIR agendas. In \textbf{Fig. 1}, we summarise the main takeaways and detailed lessons learned from five showcases. We have seen endless normative discussion on the cultural, technical, financial and educational barriers of FAIR, but little analysis exists of the structural determinants of upgrading FAIR. As such, these lessons were aggregated as shown in \textbf{Figure 1} into technical, cultural, financial and educational domain.


\begin{figure}[h]
    \centering
    \includegraphics[width=\textwidth]{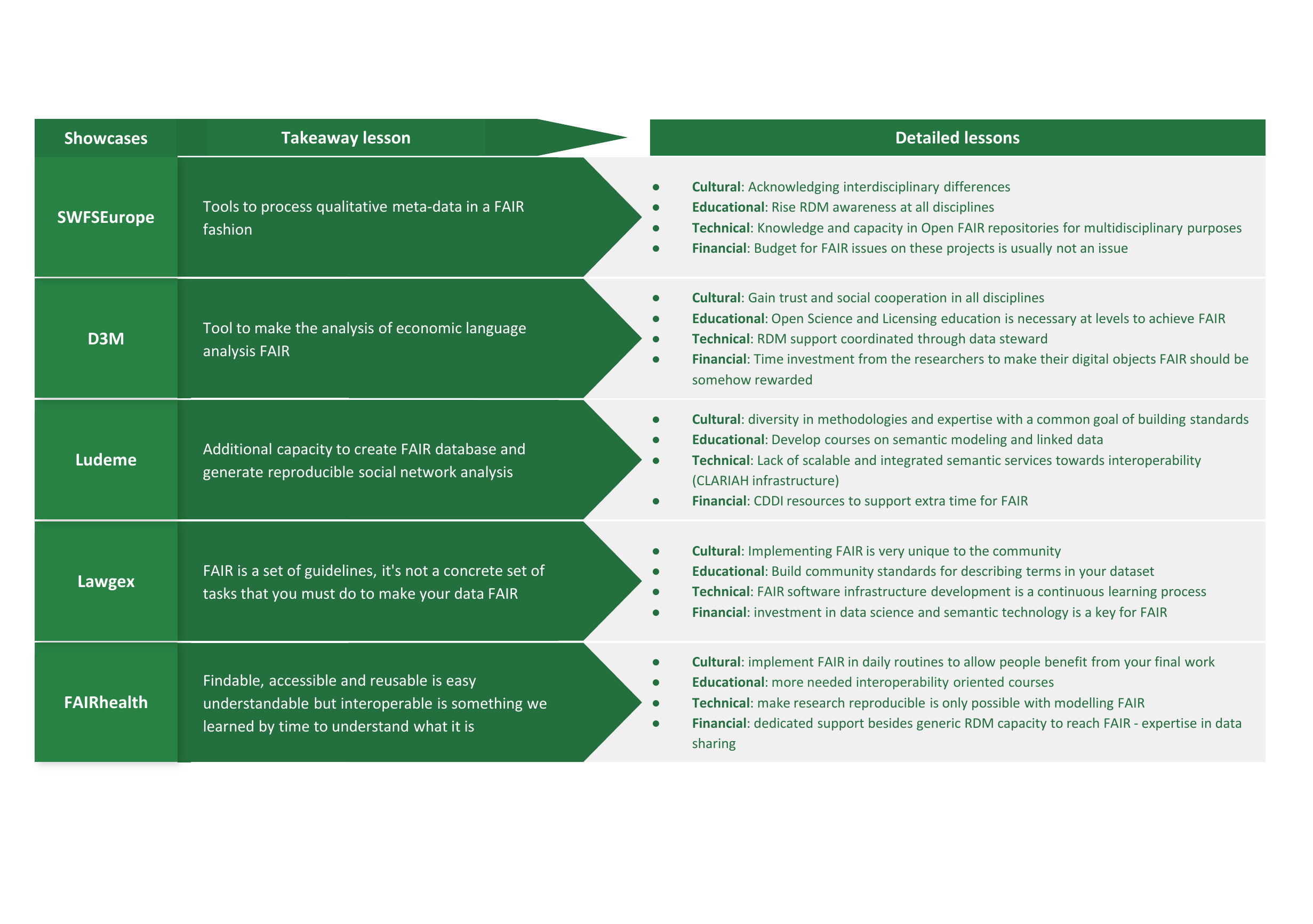}
    \caption{Overview of the main takeaways and detailed lessons per showcase}
    \label{fig:mesh1}
\end{figure}

\section{Challenges met on the road towards FAIR}

We have faced five challenges in moving UM towards universal adoption of the FAIR guidelines 1) how to best work with researchers from multiple disciplines on an array of projects; 2) how to tackle the interoperability problem 3) how to maximize our strengths as individual contributors helping researchers via the CDDI; 4) how to adopt the FAIR principles in a decentralized nature of UM; and 5) aspects of the FAIR guidelines themselves. Here we reflect on some of the creative ways that we addressed these challenges as recommendations to those who aspire to create a FAIR culture shift.

FAIR is an enabler of artificial intelligence \citep{wise2019implementation}, the intelligence with which machines, software and devices independently learn and solve problems. 

\subsection{How to best work with researchers from multiple disciplines on an array of projects}

The key challenge we found was how to provide the right answer to research data management questions and how to support researchers in their daily basis data management decisions. For instance, \textit{where to look for FAIR guidance, tools and services? which online platform can we use to share data and integrate FAIR features? or which tool help us to better automatically create metadata?}  These questions highlighted the lack of practical guidance, standard procedure and tooling collections on how to implement FAIR, in everyday research data management work. 

We noticed that FAIR research data management functions were generally perceived as a relevant and complex task in research. Yet, even foundational information about the various delicate and serious challenges that FAIR attempts to solve, remains poorly understood in many disciplines. We accepted these disciplinary differences, and constructed a work-environment to learn from each other, embarking together in the FAIR journey, involving as many stakeholders as possible. We decided to start to walk before trying to run towards FAIR: while FAIR data is an ideal we strive for, we started with a very basic support on the different showcases provided guidelines, advice on Data Management Plan creation, General Data Protection Regulation (GDPR) and tips on how to preserve the research data \footnote{Golden Rules for Good RDM: \url{https://library.maastrichtuniversity.nl/research/rdm/guide/9-golden-rules-for-good-research-data-management/}}. 

We generally reminded researchers that retrospectively 'FAIRifying' existing data is extremely resource intensive. We use 'FAIRifying' to denote the process towards achieving FAIR implementation. For all types of digital objects, we encouraged them to implement the FAIR principles as part of their research process from the very early stage of the research activity. We advised researchers that research data management (RDM) work is divided into multiple phases and requires iteratively going back and forth between different steps. Critically reflecting on each step of RDM and documenting it is mandatory for the sake of data reusability.

\subsection{How to tackle the interoperability confusion}

Significant challenge was associated with the notion of \textit{interoperability} (i.e. describing data in a machine readable format). We have found that not all researchers are motivated by the implementation of this principle, but we still encouraged them to consider how they might interact with machines when they create and consume data. One relevant question that we recommended asking is \textit{"why data has to be computably accessible by machines with user-friendly documentation using agreeing community standards?"}. Although interoperability is not the only principle, addressing these questions will ensure the strategic movement from \textit{FAR} to \textit{FAIR} University.

The level of detail of the advice provided to researchers was slightly different per showcase, given the diversity of research topics, its stage and types of digital objects. Derived from our experience, we had to provide instructions that were absolutely the most relevant and suitable for each specific stage of the RDM decision-making process. For instance, researchers in Lawgex showcase, in which experts on computer science were involved, were capable of copying concepts as semantic models and applying FAIR principles for research software (FAIR4RS) \citep{lamprecht2020towards}, while many others were struggling with the idea of using existing ontological models to turn data into machine readable format that are suitable for secondary use.

Moreover, not only researchers lacked insights into this principle. When interoperability concepts such as 'machine-actionability' were involved, it seemed to be challenging for us to provide the right support to researchers. We found that researchers are generally not aware of the technical possibilities when it comes to dealing with the application of semantics. They are very much surprised when sharing techniques that turn their data into machine-understandable format. A widely used technology solution to achieve interoperability \textit{(i.e. give context and relationship to data using community standards)} is RDF (Resource Description Framework) \citep{jacobsen2020fair}, albeit we have noticed that are not yet certainly mature in practice for all disciplines \citep{playfair}. Nevertheless, storing in institutional data repositories that support documentation such as codebooks or data dictionaries, ideally in addition to shared datasets \footnote{A guide for anyone who needs to share data: \url{https://github.com/jtleek/datasharing}}, is perhaps yet an admirable outcome of a good research data management. We kept recalling the ultimate goal as to provide as much information researchers can to help people understand and reuse your data.

\subsection{How to maximize our strengths as individual contributors helping researchers via the CDDI}

There is a crying need for services and support for research data management (\textit{what we call Research Data Service (RDS)}), but it was difficult to prioritize where to invest resources, time and budget. We found that it was complex to gain the confidence of researchers when no coordinated actions were given to deliver RDM services between staff members. We chose to prioritize integral consultative services as it would give us the opportunity to cover a basic need suitable, effective and cost-efficient. We had to not only understand researchers and their motivations toward FAIR, but most importantly, to provide central facilities to make data FAIR with the right kind of environment for collaboration, sharing of knowledge and expertise in a pragmatic way. We better organize our work when jointly shared our aspirations, goals and responsibilities individually, and work collectively as a team.

Another point of concern was the documentation delivered and maintained by the staff members. The staff tasked with providing support should be responsible for producing a solid, realistic and daily-based action plan for supporting researchers and for the implementation of a university-wide FAIR. The documentation of their related activities, FAIR solutions, best practices, and learning should be reusable, trackable, and easily accessible to prevent the loss of important information due to organizational or technical changes. It also ensures reuse of materials and solutions that already exist, rather than starting from scratch \citep{garcia2020ten}.



We also observed that we should particularly have a keen interest in actively participating in formal or informal conversations, and community channels between researchers about data standards for interoperability and reusability, ensuring they are accepted by as many individuals as possible. Researchers are advised to join their community to change the culture of data management and data sharing, and contribute on how FAIR technology is implemented in the ways community standards meet, or fail to meet.

A sustainable part of our job has been raising awareness about the importance of suitable RDM (i.g.training skills for managing, storing and exploiting data)\footnote{ FAIR essential workshop (lecture and hands on exercises): \url{https://github.com/MaastrichtU-IDS/fair-workshop}}. In addition to essential workshops about FAIR, other training for applying semantics using digital techniques, software best practices \footnote{Best practices and documentation at the IDS: \url{https://maastrichtu-ids.github.io/best-practices/}} or tools for reproducible research software were required as interoperability remains a central problem in many disciplines. When in theory training 50 researchers could have made a more impactful than supporting on a single showcase, it was often hard in practice to find the time for teaching and developing the course as we were so busy managing our FAIR obligations.\citep{software}.

\subsection{The structure of the university (or other organisation)}

We observed that the structure of the organisation must be taken into account when thinking about implementing FAIR. At our university, a top-down approach for implementing FAIR would not have been effective because of the decentralized nature of the institution. Each faculty (what North Americans would recognize as a “college” or “school”) sets its own policies regarding expectations for researcher behavior, though some research data and computing support services such as the CDDI and the Data Science Research Infrastructure (DSRI) are available in a centralized manner. Following the FAIR principles to store data is a key, but it is not enough by itself. At UM, a diverse working group (central library, ICT, representative of all faculties) on RDM and computing services was established to ensure the guidelines included the diversity of needs.

We live in an information age where we spend increasing amounts of money on activities about which we have little solid information, a system to check and evaluate the status of our ambition to become a FAIR university. In this context it would be useful to have a complete dashboard with an array of indicators to track what makes FAIR investment worthwhile: community service, jobs, knowledge and disciplinary cohesion. Despite the difficulty of quantifying these things, they are needed for guiding our actions towards FAIR. The current unquantifiable nature of FAIR impact on researchers’ projects makes work more difficult, and it may also undermine the ability to trust FAIR in general.  

FAIR is at an early stage, and it involves a learning process between stakeholders. Researchers and the staff members should come together as part of a strong community with shared goals and accountability \citep{staff}. Establishing an engaged network of data scientists (and stewards) across departments requires commitment from the different actors and can be enabled through regular meetings and social events, which facilitate knowledge sharing, create a sense of community with a common purpose, and foster collaborations across departments. “Leaders” of the different departments can facilitate this by forming a governance body that promotes such initiatives and implements a common strategy based on cross-departmental collaborations, university-wide data science technological platform, and collaborative investment in talent development.

It is also important to keep in mind the working conditions of employees. Achieving FAIR in its fullest form is a long-term vision, not something that can be realized immediately. This was relevant when hiring and evaluating data stewards, data scientists, and other data experts, so that appropriate and realistic goals are set. People in these roles are on the front lines of culture change -- no policy change will give them the power to achieve what they want. As a result, they all need to engage and explore proactively opportunities with the help of a strategic figure such as the data steward, who must be somewhat familiar with domain-specific digital tools and methods.

\subsection{The nature of the FAIR guidelines themselves}

One strength of the FAIR guidelines is their implicit openness to knowledge diversities; there is a range of ways to implement FAIR, all of which are valid. This strength is a double-edged sword, however: because these are guidelines, not standards, the requirements are broad. Another strength of the guidelines as they are written is their precision, but on the other hand they do not specify which humans in organisations ought to be responsible for implementing certain things. These are irresolvable challenges for which every university must find its own solution. 

Another challenge is the theory of assuming the FAIR guidelines must be followed and implemented in order (e.g., first “F,” then “A”), when in practice the FAIR implementation is rather an iterative and dynamic process. We suggested several times identifying which principles are taking place at different stages of the RDM lifecycle or Data Science workflow (e.g. Mapping FAIR principles to the data Data Science Life Cycle or the generic workflow proposed \citep{workflow}).

As much as we found challenging implementing the principles, we found it useful to debate the relationship and potential barriers or drivers between FAIR and other ethical and responsible RDM frameworks as CARE\citep{carroll2020care} and TRUST \citep{lin2020trust}. A deeply understanding and practical clarification between FAIR and already established technologies and approaches such as Linked Data or semantic technologies should be provided as to potentially open new inter-faculty collaborations.

\section{Future directions}

Many challenges have been addressed and faced in the CDDI. Work is currently focused on new solutions for scaling up the  general adoption of FAIR and new ways to put them in practice. We briefly describe several preliminary ideas for future work and our plans (see \textbf{Figure 3}):

\begin{figure}[h]
  \centering
  \includegraphics[width=\linewidth]{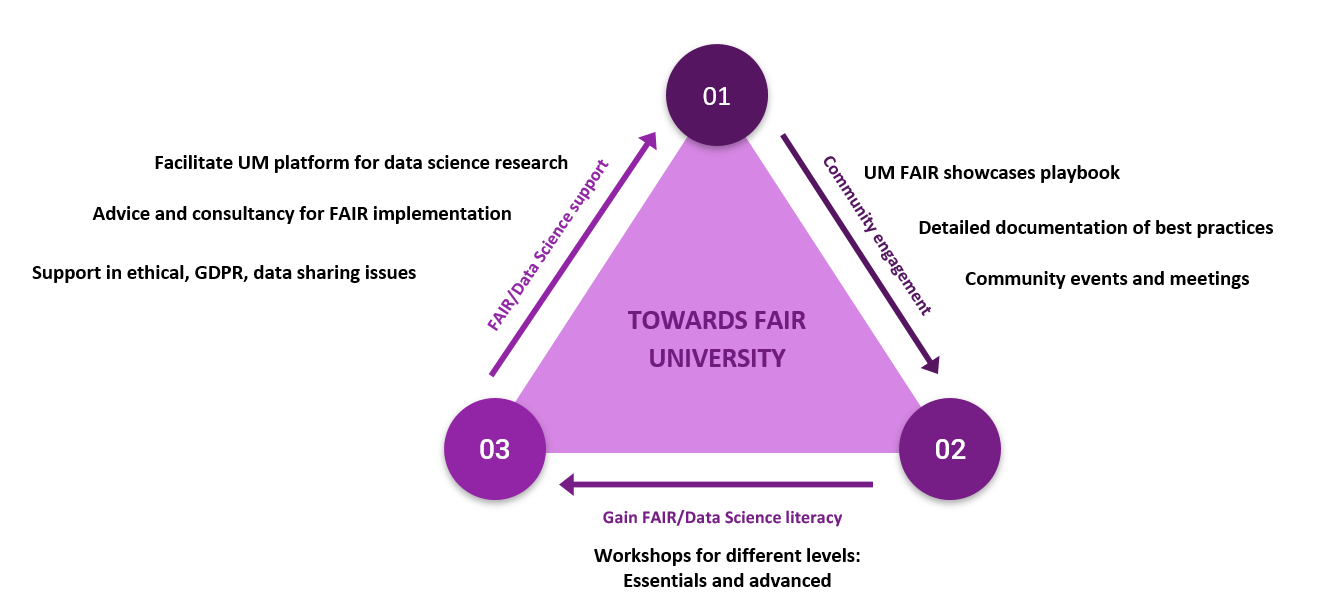}
  \caption{Overview of the actions plan towards FAIR. \citep{pilots} [Public domain], (\url{https://doi.org/10.6084/m9.figshare.14790102.v2}).}
\end{figure}

\begin{itemize}
  \item  \textit{1. Community engagement:} Understanding stage-specific barriers and drivers of FAIR adoption, develop tailored FAIR courses. \footnote{Link to UM services, tools and training for RDM: \url{https://library.maastrichtuniversity.nl/research/rdm/services-tools-training/}} and FAIR Workshops.
  \item \textit{2. FAIR and Data Science literacy:} Train researchers at different levels, not only essentials about FAIR principles but also hands-on examples on how to build open-science software and technology for interoperability such as data conversion using public services. 
  \item \textit{3. FAIR and Data Science support:} Facilitate a university-wide platform for data science as well as give advanced support and consultancy on FAIR practical implementation and GDPR.
\end{itemize}

\section{Conclusion}
The UM CDDI lessons tell us a story about how FAIR could be achieved at an institutional level, as long as there are resources that can support diverse projects in distinctive fields. These resources have to accommodate a variety of epistemological values, and have the flexibility to be deployed in various types of expertise where it is needed.The FAIRification of an entire university is a long game, but important short-term gains with ongoing basis individual projects can help to inspire other researchers to begin their own FAIR journey.  Although we are FAR - we acknowledge the limited actions towards interoperability -  from our objective, we need to aim to become a FAIR University. 

\section{Acknowledgments}
The authors would like to acknowledge the UM for having initiated the CDDI project where this report started. Special thanks to the CDDI past and current members, for giving us the opportunity for having real experience on FAIR data management, and in particular Ricardo Moreira da Rocha and Lianna Ippel. Also special mention to Michael Erard who guided us through this writing process.

\bibliographystyle{unsrt}  
\bibliography{references}






\end{document}